\documentclass[preprints,article,accept,moreauthors,10pt,a4paper]{Definitions/mdpi} 

\firstpage{1} 
\pubvolume{xx}
\issuenum{1}
\articlenumber{1}
\pubyear{2018}
\copyrightyear{2018}
\externaleditor{Academic Editor: name}
\history{Received: date; Accepted: date; Published: date}

\usepackage{epsfig}
\usepackage{amsmath}
\usepackage{amsfonts}
\usepackage{amssymb}
\usepackage{graphicx}
\usepackage{colordvi}
\usepackage{color}
\usepackage[alsoload=astronomy]{siunitx}

\newcommand{\aea}{Astron. Astrophys.}
\newcommand{\apj}{Astrophys. J.}
\newcommand{\apjs}{Astrophys. J. Suppl.}
\newcommand{\apjl}{Astrophys. J. Lett.}

\newcommand{\grg}{Gen.  Rel. Grav.}
\newcommand{\jcap}{J. Cosmol. Astropart. Phys.}
\newcommand{\mnras}{Mon. Not. R. Astron. Soc.}
\newcommand{\prd}{Phys. Rev. D}

\newcommand{\plb}{Phys. Lett. B}

\Title{Decaying dark energy in light of the latest cosmological dataset}

\Author{Ivan de Martino $^{1}$\orcidA{}}

\AuthorNames{Ivan de Martino}

\address{%
$^{1}$ \quad Department of Theoretical Physics and History of Science, University of the Basque Country UPV/EHU,
Faculty of Science and Technology, Barrio Sarriena s/n, 48940 Leioa, Spain; ivan.demartin@ehu.eus}

\corres{Correspondence: ivan.demartin@ehu.eus}

\abstract{ Decaying Dark Energy models modify the background evolution of the most common observables, such as the Hubble function, the luminosity distance and the Cosmic Microwave Background  temperature-redshift scaling relation. We use the most recent observationally-determined datasets, including Supernovae Type Ia and Gamma Ray Bursts data, along with $H(z)$  and Cosmic Microwave Background  temperature versus $z$ data and the reduced Cosmic Microwave Background parameters,  to improve the previous constraints on these models. We perform a Monte Carlo Markov Chain analysis to constrain the parameter space, on the basis of two distinct methods. In view of the first method, the Hubble constant and the matter density are left to vary freely. In this case, our results are compatible with previous analyses associated with decaying Dark Energy models, as well as with the most recent description of the cosmological background. In view of the second method, we set the Hubble constant and the matter density to their best fit values obtained by the {\it Planck} satellite, reducing the parameter space to two dimensions, and improving the existent constraints on the model's parameters. Our results suggest that the accelerated expansion of the Universe is well described by the cosmological constant, and we argue that forthcoming observations will play a determinant role to constrain/rule out decaying Dark Energy.}

\keyword{Dark Energy; Statistical analysis; BAO; Supernovae; Cosmological model; Hubble constant; CMB temperature.}

\begin{document}

\section{Introduction}

In the last decades, several observations have pointed out that the Universe is ongoing a period of accelerated expansion that is driven by the presence of an exotic fluid with negative pressure \cite{Planck18,Perlmutter1997,Riess2004,Astier2006,Suzuki2012,Pope2004,Percival2001,Tegmark2004,Hinshaw2013,Blake2011,demartino2015b,demartino2016b}. Its simplest form is a cosmological constant $\Lambda$, having an equation of state $w=-1$. More complicated prescriptions lead to the so-called Dark Energy (DE).
Although several models have been proposed to explain DE \cite{PR03,Pad03,D+05,CTTC06,Caldwell02,PR88,RP88,SS00,Capolupo2016,Capolupo2017,Kleidis2011,Kleidis2011b,Kleidis2015,Kleidis2016,Kleidis2017}, the observations have only determined that it accounts for $\sim 68\%$ of the total energy-density budget of the Universe, while its fundamental nature is still unknown (see for instance the reviews \cite{caldwell, atrio2016}). Also, we should mention that the accelerated expansion of the Universe could be explained by several modifications of the gravitational action. For example, introducing higher order terms of the Ricci curvature  in the Hilbert-Einstein Lagrangian, gives rise to an effective matter stress-energy tensor which could drive the current accelerated expansion (see for example the following reviews \cite{darkmetric, Nojiri2011, idm2015, Nojiri2017, Nojiri:2006ri, PhysRept, demartino2016b}). Another alternative for reproducing the dark energy effects is by introducing non-derivative terms interactions in the action, in addition to the Einstein-Hilbert action term, such that it creates the effect of a massive graviton \cite{Arraut2015,Arraut2016,Arraut2017}.

We are interested into explore a specific decaying DE model, $\Lambda (z) \propto (1+z)^{m}$, leading to creation/annihilation of photons and Dark Matter (DM) particles. The model is based on the theoretical framework developed in 
\cite{lima+96,LA99,lima+00,puy,ma}, while the thermodynamic features have been developed in \cite{jetzer10,jetzer11}. Since DE continuously decays into photons and/or DM particles along the cosmic evolution, the relation between the temperature of the Cosmic Microwave Background (CMB) radiation and the redshift is modified.

In the framework of the standard cosmological model, the Universe expands adiabatically and, as consequence of the entropy and photon number conservation, the temperature of the CMB radiation scales linearly with redshift, $\propto(1+z)$. Nevertheless, in those models where conservation laws are violated, the creation or annihilation of photons can lead to distortions in the blackbody spectrum of the CMB and, consequently, to deviations of the standard CMB temperature-redshift scaling relation. Such deviations are usually explored with a phenomenological parametrization, such as $T_{\rm CMB}(z) = T_0 (1+z)^{1-\beta}$ proposed in \cite{lima+00}, where $\beta$ is a constant parameter ($\beta=0$ means adiabatic evolution), and $T_0$ is the CMB temperature at $z=0$, which has been strongly constrained with COBE-FIRAS experiment, $T_0=2.7260 \pm 0.0013$~K \cite{fixsen09}.  The parameter $\beta$ has been constrained using two methodologies: (a) the fine structure lines corresponding to the transition energies of atoms or molecules, present in quasar spectra, and  excited by the CMB photon \cite{bahcall68}; (b) the multi-frequency  measurements of the Sunyaev-Zel'dovich (SZ) effect \cite{sunyaev70, fabbri78,rephaeli80}. Recent results based on data released by the {\it Planck} satellite and the {\it South Pole Telescope} (SPT) have led to sub-percent constraints on $\beta$ which results to be compatible with zero at $1\sigma$ level (more details can be found in \cite{demartino12,demartino2015b,luzzi09,luzzi15,hurier14,saro14,Avgoustidis2016}).

In this paper, we start with the theoretical results obtained in \cite{jetzer10,jetzer11}. 
Such a model has been constrained using luminosity distance measurements from Supernovae Type Ia (SNIa), differential age data, Baryonic Acoustic Oscillation (BAO), the CMB temperature-redshift relation, and the CMB  shift parameter. Since the latter depends on the redshift of the last surface scattering, $z_{CMB}\sim 1000$, it represents a very high redshift probe. On the contrary, other dataset were used to probe the Universe at low redshift, $z\lesssim 3.0$. We aim to improve those constraints performing two different analysis: in the first one, we will constrain the whole parameter space to study the possibility of  the model to alleviate  the tension in the Hubble constant (see Sect. 5.4 in \cite{Planck18} for the latest results on the subject); in the second analysis, we will adopt the {\it Planck} cosmology  to improve the constraint on the remaining parameters.
Thus, we retain the SNIa, and use the most recent measurements the differential age, BAO, and the CMB temperature-redshift data. 
Also, we use luminosity distances data of Gamma Ray Burst (GRB), which allow us to extend the redshift range till $z\sim 8$. 
Finally, we will also use  the reduced (compressed) set of parameters from CMB constraints \cite{Planck18}. 

The paper is organised as follows. In Sect. \ref{sec:theory} we summarize the theoretical framework starting from the general Friedman-Robertson-Walker (FRW) metric, and pointing out the modification to the cosmological background arising from the violation of the conservation laws. In Sect. \ref{sec:data} we present the datasets used in the analysis, and the methodology implemented to explore the parameter space. The results are shown and discussed in Sect. \ref{sec:results} and, finally, in Sect.
\ref{sec:conclusions} we give our conclusions.

\section{Theoretical framework}\label{sec:theory}

The starting point is the well-known FRW metric
\begin{equation}
ds^2 = c^2dt^2 -a^2(t)\biggl[\frac{dr^2}{1-kr^2} + r^2(d\theta^2+\sin^2\theta d\phi^2)\biggr]\,,
\end{equation}
where $a(t)$ is the scale factor and $k$ is the curvature of the space time \cite{Weinberg72}. In General Relativity (GR), one obtains the following Friedman equations:
\begin{align}
&8\pi G(\rho_{m,tot}+\rho_x)+ \Lambda_0c^2= 3\biggl(\frac{\dot a}{a}\biggr)^2+3
\frac{kc^2}{a^2}~, \label{eq:Fried1}\\
&\frac{8\pi G}{c^2}(p_{m,tot}+p_x)- \Lambda_0c^2= -2\frac{\ddot a}{a}- \frac{\dot
	a^2}{a^2}-\frac{kc^2}{a^2}~\,, \label{eq:Fried2}
\end{align}
where the total pressure is  $p_{m,tot} = p_{\gamma}$, the total density is $\rho_{m,tot} = \rho_{m} +
\rho_{\gamma}$, and $\rho_x$ and $p_x$ are the density and pressure of DE, respectively. Following \cite{jetzer10,jetzer11}, we set both the 'bare' cosmological constant $\Lambda_0$ and the curvature $k$ equal to 0. 

In the standard cosmology, the Bianchi identities hold and the stress-energy momentum $T^{\mu \nu}$ is locally conserved 
\begin{equation}
\nabla_{\mu} T^{\mu \nu} =0\,.
\end{equation}
Adopting a perfect fluid, the previous relation can recast as
\begin{equation}
\dot\rho +3 (\rho+p) H = 0  ~,\label{eq:cons_ALL}
\end{equation}
where $H \equiv \dot a / a$ is the definition of the Hubble parameter. Thus, each component is conserved. Nevertheless, due to the photon/matter creation/annihilation happening in the case of decaying DE, the conservation equation is recast in the following relations:
\begin{align}
& \dot\rho_{m} +3 \rho_{m} H = (1-\epsilon) ~ C_{x}
~,\label{eq:cons_mat}\\
&\dot\rho_{\gamma} +3 \gamma \rho_{\gamma} H = \epsilon ~ C_{x}~,\label{eq:cons_gamma}\\
&\dot\rho_{x} +3(p_{x} + \rho_{x})H = - C_{x} ~,\label{eq:cons_x}
\end{align}
where $\gamma$ is a free parameter determining the equation of state of radiation $p_\gamma=(\gamma-1)\rho_\gamma$ and, $C_{x}$ and $\epsilon$ account for the decay of DE. $C_x$ describe the physical mechanism leading to 
the production of particles  (see, for instance, the thermogravitational quantum creation theory \cite{LA99} or the quintessence scalar field cosmology \cite{RP88}), and $\epsilon$ must be small enough in order to have the current density of radiation matching the observational constraints.
Assuming $p_x=-\rho_x$, and defining
\begin{equation}
\rho_x=\frac{\Lambda(t)}{8\pi G}\,,
\end{equation}
the parameter $C_x$ can be obtained from the Eq. \eqref{eq:cons_x}
\begin{equation}
C_x= -\frac{\dot\Lambda(t)}{8\pi G}\,. \label{eq:Cx}
\end{equation}

Following \cite{jetzer10,jetzer11}, one can adopt a power law model
\begin{equation}
\Lambda(t) = B \biggl(\frac{a(t)}{a(0)}\biggr)^{-m} =  B (1+z)^{m}\,,
\end{equation}
then, writing  Eq. (\ref{eq:Fried1}) at the present epoch, one can obtain $B=3 H_{0}^{2} (1- \Omega_{m,0})$, where $\Omega_{m,0}$ is the matter density fraction at $z=0$. It is very straightforward to verify that setting the power law index $m=0$ leads to the cosmological constant. From Eq. \eqref{eq:cons_x}, it is also possible to write down an effective equation of state for the DE \cite{jetzer11}:
\begin{equation}\label{eq:weff}
w_{eff} = \frac{m}{3}-1\,.
\end{equation}

Finally, using Eqs. \eqref{eq:Fried1}, \eqref{eq:cons_mat}, \eqref{eq:cons_gamma}, and \eqref{eq:cons_x}, the Hubble parameter can be obtained \cite{ma,jetzer10,jetzer11}:
\begin{align}\label{eq:H}
H(z) & \simeq  \frac{8 \pi G}{3} (\rho_{m} + \rho_{x})
 =  H_0 \left[\frac{3 (1-\Omega_{m,0})}{3 - m} (1+z)^{m} +
\frac{(3\Omega_{m,0}- m) }{3-m} (1+z)^{3} \right]^{1/2}\,.
\end{align}

Let us note that  the standard Hubble parameter is recovered by setting $m=0$ in Eq. \eqref{eq:H}. 
Having the Hubble parameter, allow us to compute the luminosity distance as follows
\begin{equation}\label{eq:dl}
	D_L =\frac{(1+z)c}{H_0}\int_{0}^{z}\frac{dz'}{E(z')}\,,
\end{equation}
where we have defined $E(z)\equiv H(z)/H_0$.

Finally, following the approach originally proposed in  \cite{lima+96}, combining the Eqs. \eqref{eq:cons_mat}, \eqref{eq:cons_gamma}, and \eqref{eq:cons_x}, with the equation for the number density conservation
\begin{align}
& \dot n_\gamma + 3n_\gamma H= \psi_\gamma~, \label{nn}
\end{align}
where $\psi_\gamma$ is the photon source, and with the Gibbs Law
\begin{align}
&  n_\gamma T_\gamma d\sigma_\gamma = d\rho_\gamma - \frac{\rho_\gamma +p_\gamma }{n_\gamma}dn_\gamma\,, 
\label{gibbs}
\end{align}
one obtains,  through the use of thermodynamic identities, the following CMB temperature redshift relation (see for more details \cite{ma,jetzer11}): 
\begin{align}\label{ma3}
T_{CMB}(z) & = T_0 (1+z)^{3(\gamma-1)} \times
\left(\frac{(m-3\Omega_{m,0})+m(1+z)^{m-3}(\Omega_{m,0}-1)}{(m-3)\Omega_{m,0}}\right)^{(\gamma-1)}~.
\end{align}

\begin{figure}[!ht]
	\centering
	\includegraphics[width=7.5 cm]{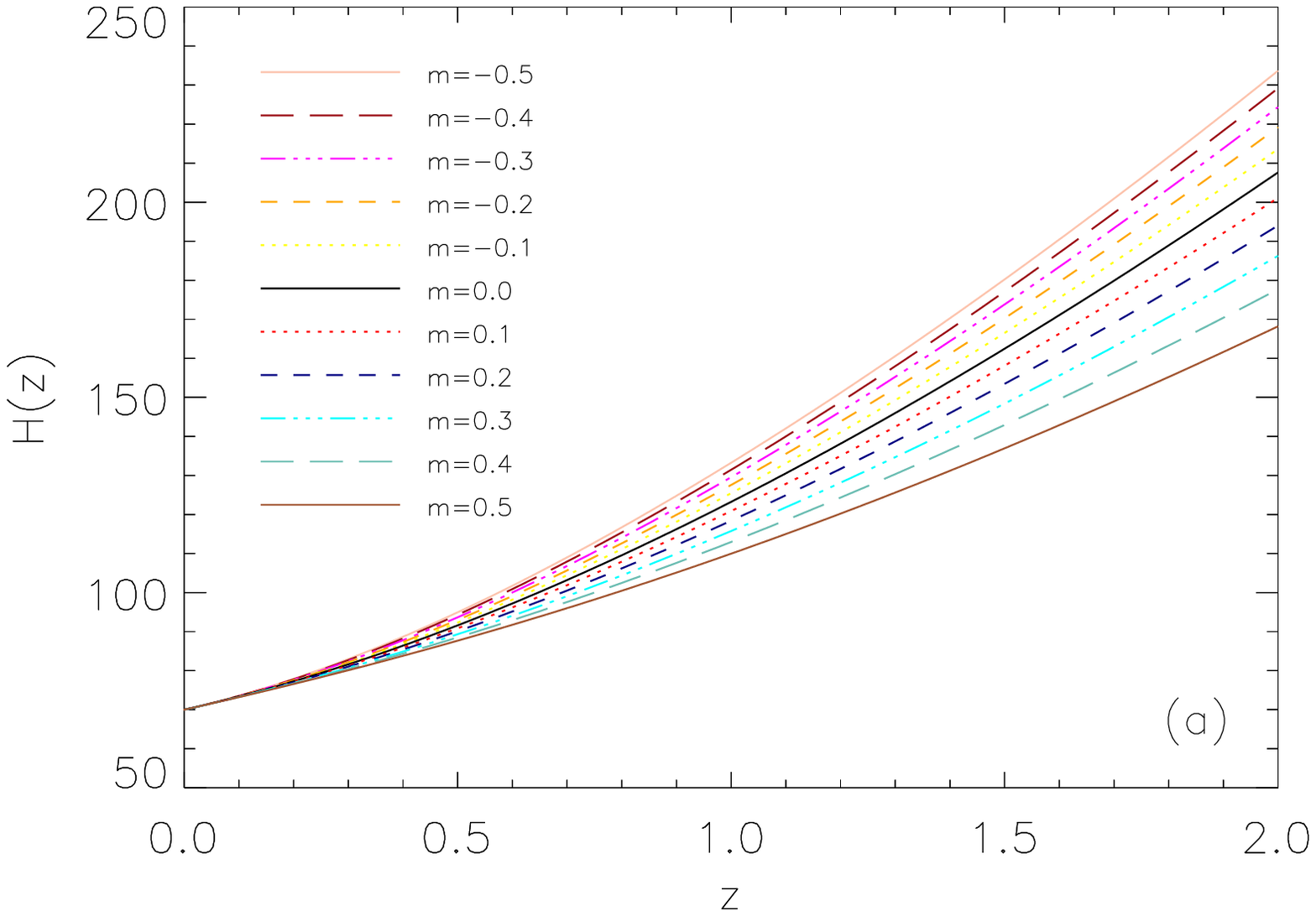}
	\includegraphics[width=7.5 cm]{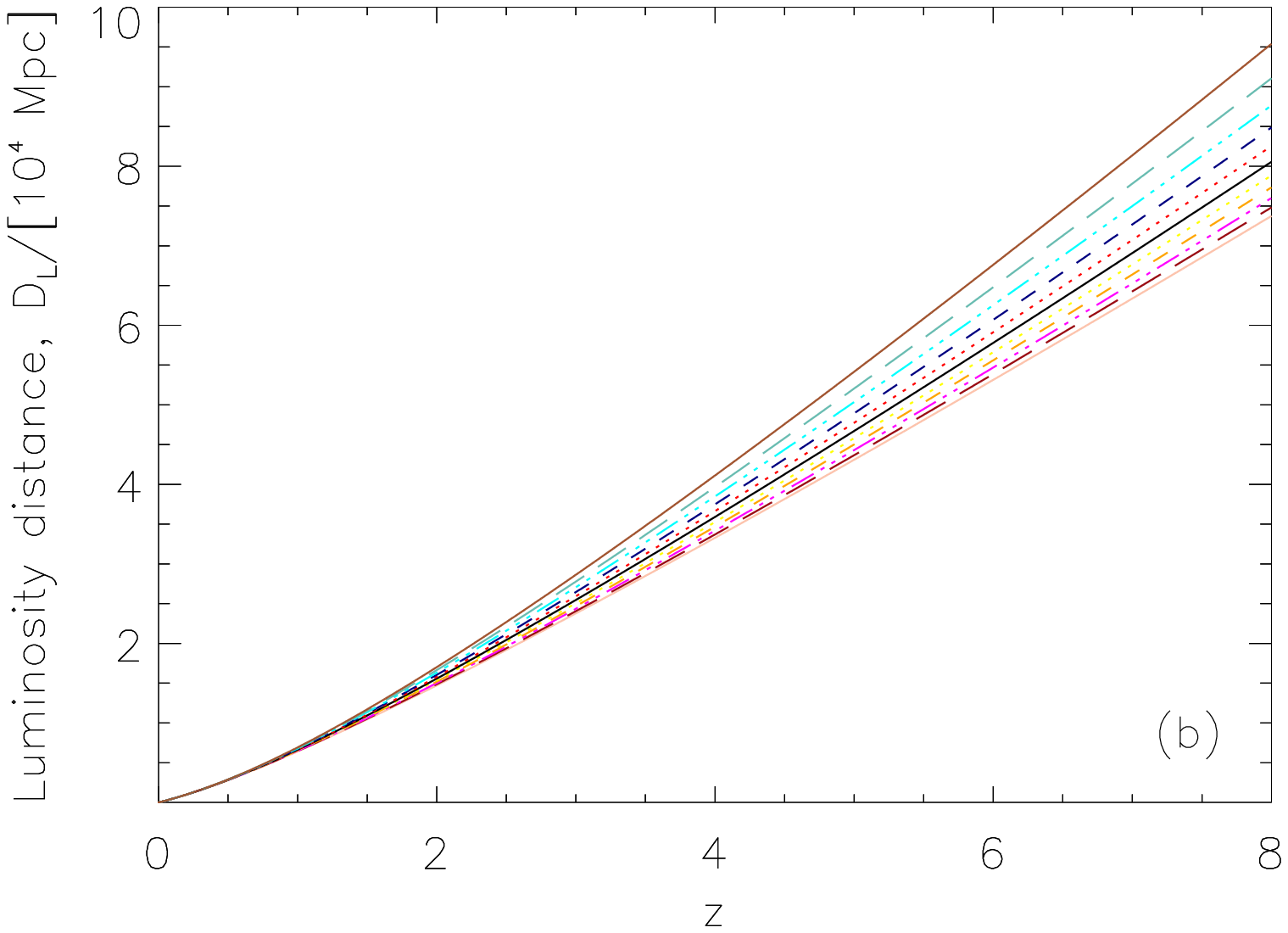}\\
	\includegraphics[width=7.5 cm]{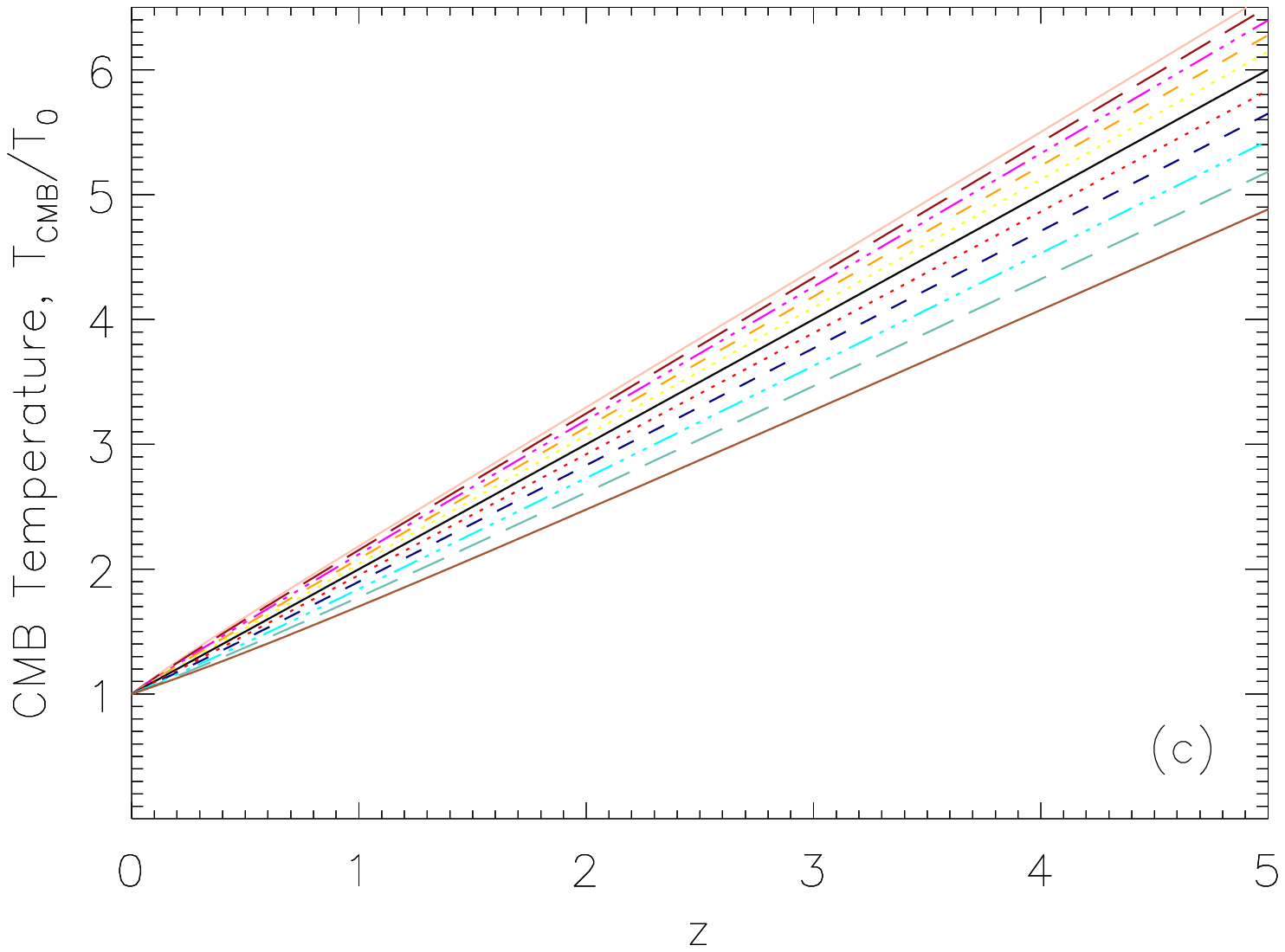}
	\includegraphics[width=7.5 cm]{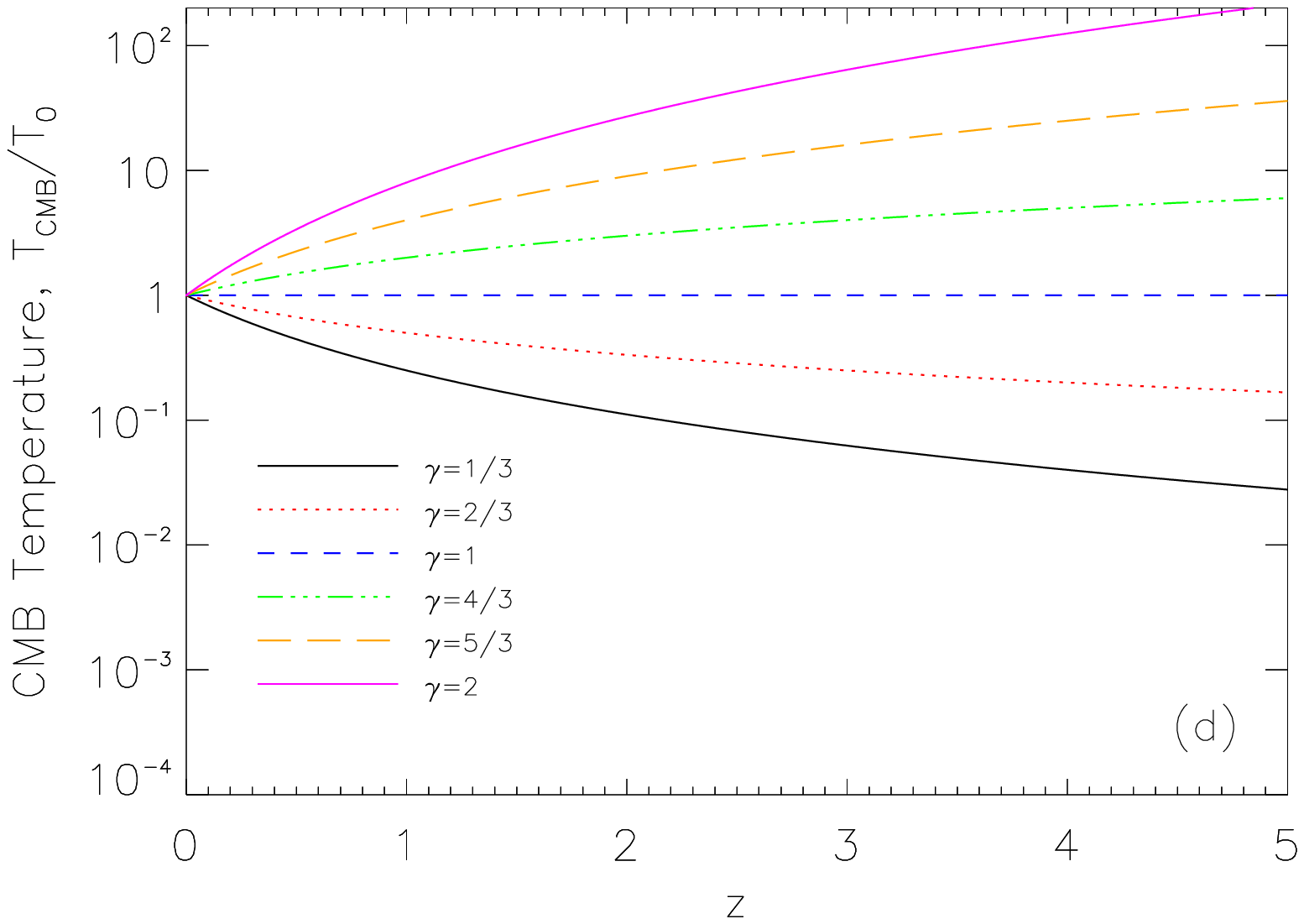}\\	
	\caption{The figure shows the Hubble constant, the luminosity distance and the CMB temperature as function of redshift. Colors and lines indicate the different values assigned to the parameters $m$ and $\gamma$ in order to illustrate their impact on the observables.}\label{fig1}
\end{figure}

Again, setting $m=0$ gives the standard relation $T_{CMB}(z)=T_0(1+z)$.
Eqs. \eqref{eq:H}, \eqref{eq:dl} and \eqref{ma3} can be easily implemented to test the decaying DE scenario. To show the effectiveness of these observables in constraining the cosmological model, we depict in Fig. \ref{fig1} their scalings as a function of the redshift for different value of the parameters $\gamma$  and $m$, while we set $H_0$ and $\Omega_0$ to their best fit from {\it Planck}
satellite. In the panels (a), (b), and (c) we fix
$\gamma=4/3$ (which represents its standard value) while varying $m$ in the range $[-0.5, 0.5]$ to show its impact on the Hubble constant, the luminosity distance and the CMB temperature. On the contrary, in panel (d) we set $m=0$ (standard value) and vary $\gamma$ illustrating how much the  $T_{CMB}$-redshift relation is affected.  The redshift ranges in the panels are set to the ones of the datasets. Looking at the plots, it is clear that the data will be really sensible to a variation of $\gamma$, while $m$ will be more difficult to constrain.

\section{Methodology and data}\label{sec:data}

We use measurements of $H(z)$, of luminosity distances from SNIa and GRBs, of BAO, and of the CMB temperature redshift relation. Then, we predict the theoretical counterparts using Eqs. \eqref{eq:H}, \eqref{eq:dl}, and \eqref{ma3}, and we fit each one to the corresponding dataset computing the likelihood $-2\log{\cal L}=\chi^2({ \bf p})$, where 
 ${\bf p}=[H_0, \Omega_{m,0}, m, \gamma]$ are 
the parameters of the model. 
The parameter space is explored  using a Monte Carlo Markov Chain (MCMC) based on the Metropolis-Hastings \cite{Metropolis1953, Hastings1970} sampling  algorithm with an adaptive step size to guarantee an optimal acceptance rate between 20\% and 50\% \cite{Gelman1996, Roberts1997}, while the convergence is ensured by the Gelman-Rubin criteria  \cite{Gelman1992}.  Once the convergence criteria is satisfied,  
the different chains are merged to compute the marginalized likelihood  ${\cal L}({\bf p})=\Pi_k{\cal L}({\bf p})$, where $k$ indicates the different datasets, and to constrain the model's parameters.
The priors are specified in Table \ref{tab:priors}. 
\begin{table}[!h]
	\begin{center}
		\begin{tabular}{|cc|}
			\hline
			{\bf Parameter} & {\bf Priors}\\
			\hline
			$H_0$  & $[50.0, 100.0]$ \\
			$\Omega_{m,0}$ & $[0.0, 1.0]$\\
			$m$ & [-1, 1]\\
			$\gamma$ & $[1.0, 2.0]$  \\
			\hline
		\end{tabular}
		\caption{Parameter space explored by the MCMC algorithm.}\label{tab:priors}
	\end{center}
\end{table}
Finally, the expectation value ($\langle {p}_i\rangle$) of the 1D marginalized likelihood distribution and the corresponding variance 
are computed as follows \cite{Spergel2003}
\begin{align}
& \langle {p}_i\rangle=\int d^{N_s}{\bf p} \mathcal{L}({\bf p}) {p}_i,\label{eq:bestfit}\\
&  \sigma^2_i=\int d^{N_s}{\bf p}  \mathcal{L}({\bf p}) ({p}_i -  \langle{p}_i\rangle)^2\label{eq:errorbestfit},
\end{align}
where  $N_s$ is the dimension of the parameter space. 

Finally, the joint likelihood of the independent observables will be used to compare decaying DE model with $\Lambda$CDM employing the Akaike Information Criteria (AIC) \cite{Akaike1974}:
\begin{equation}
AIC = -2\log\mathcal{L}_{max} + 2N_p\,,
\end{equation}
 where $N_p$ is the number of parameters. A negative variation of the AIC indicator with respect to the reference model, $\Delta(AIC)=AIC_{dec.DE}-AIC_{\Lambda CDM}$, would indicate the model performs better than $\Lambda CDM$.
 
\subsection{Supernovae Type Ia}\label{sec:SN}

We use a dataset  of 557  Supernovae Type Ia  (SNIa) in the redshift range $z=[0,1.4]$
extracted from the UnionII catalogue  ( more details can be found in \cite{Amanullah+08}). The observable is the so-called distance modulus $\mu_{obs}$, which is the difference of the apparent and absolute magnitudes. Its theoretical counterpart can be computed starting from the luminosity distance in Eq. \eqref{eq:dl}, and it is given by
\begin{equation}\label{eq:distmod}
\mu_{th} (z) = 5 \log_{10}\hat{D}_{L}(z) + \mu_{0}~,
\end{equation}
where $\mu_{0} = 42.38 - 5 \log_{10}h$, with $h\equiv H_0/100$, and $\hat{D}_{L}(z)$ is given by
\begin{equation}
\hat{D}_{L}(z) = (1+z)\int_{0}^{z}\frac{dz'}{E(z')}.
\end{equation}

Then, we can define the $\chi^{2}$ function as
\begin{equation}
-2\log{\cal L}_{\mathrm{SN}}({ \bf p})=\chi_{\mathrm{SN}}^{2}({ \bf p})=\sum_{i=1}^{557}\biggl(\frac{\mu
	_{th}(z_{i}, { \bf p})-\mu _{obs}(z_i)}{\sigma _{\mu}(z_i)}\biggr)^{2},
\label{eq:chiSN}
\end{equation}
where $\sigma_{\mu}(z)$ is the error on $\mu _{obs}(z)$.
Let us note that the parameter $\mu_{0}$ encodes the dependence by the Hubble constant. Whenever one is not interest in fitting $H_0$, 
the marginalized  $\chi^{2}$ function can be defined as 
\cite{DiPietro+03, Nesseris+05,	Perivolaropoulos05, Wei10a}:
\begin{equation}
\tilde{\chi} _{\mathrm{SN}}^{2}({ \bf p})= \tilde{A} -
\frac{\tilde{B}^{2}}{\tilde{C}}, \label{eq:chiSN_2}
\end{equation}
where
\begin{align}
\tilde{A} & = \sum_{i=1}^{557}\biggl(\frac{\mu _{th}(z_{i},{ \bf p}, \mu_{0} =
	0)- \mu _{obs}(z_i)}{\sigma _{\mu}(z_i)}\biggr)^{2}, \\
\tilde{B} & = \sum_{i=1}^{557}\frac{\mu _{th}(z_{i},{ \bf p}, \mu_{0} =
	0)- \mu _{obs}(z_i)}{\sigma _{\mu}^{2}(z_i)}~, \\
\tilde{C} & = \sum_{i=1}^{557}\frac{1}{\sigma _{\mu }^{2}(z_i)}~.
\end{align}

\subsection{Differential ages, $H(z)$}

Following \cite{Lukovic2016}, we use 30 uncorrelated measurements of expansion rate, $H(z)$, that have been obtained using the differential age method \cite{Jimenez02,Simon05, Stern10, Moresco12a,Moresco12b, Moresco16a, Moresco15, Zhang14}. Thus, we define the corresponding $\chi ^{2}$ as
\begin{equation}
-2\log{\cal L}_{\mathrm{H}}({ \bf p})=\chi _{\mathrm{H}}^{2}({ \bf p})=\sum_{i=1}^{30}\biggl(\frac{H(z_{i},{ \bf p})-
	H_{obs,}(z_{i})}{\sigma _{H}(z_{i})}\biggr)^{2}~, \label{eq:chiOHD}
\end{equation}
where $\sigma_{H}(z)$ is the error on $H_{obs}(z)$. As stated in  Sec. \ref{sec:SN}, 
the marginalized $\chi ^{2}$ with respect to $H_{0}$ can be also defined using Eq. (\ref{eq:chiSN_2}), where for the $H(z)$ dataset we have
\begin{align}
\tilde{A} & = \sum_{i=1}^{30}\biggl(\frac{(H(z_{i},{ \bf p}, H_{0}=1)-
	H_{obs}(z_i)}{\sigma _{H}(z_i)}\biggr)^{2}, \\
\tilde{B} & = \sum_{i=1}^{30}\frac{H(z_{i},{ \bf p},H_{0}=1)-
H_{obs}(z_i)}{\sigma _{H}^{2}(z_i)}~, \\
\tilde{C} & = \sum_{i=1}^{30}\frac{1}{\sigma _{H}^{2}(z_i)}~.
\end{align} 

\subsection{Baryonic Acoustic Oscillation}  

It is customary to define the BAO's observable as the following ratio: $\hat\Xi\equiv r_d/D_V(z)$; where  $r_{d}$ is the sound horizon at the drag epoch $z_d$ \cite{Eisenstein98}:
\begin{equation}
r_d =\int_{z_d}^{\infty} \frac{c_s(z)}{H(z)}dz\,,
\label{eq:r_d}
\end{equation}
and $D_V$ the spherically averaged distance measure \cite{Eisenstein05}
\begin{equation}
D_V(z)\equiv \left[(1+z)^2 d_A^2(z)\dfrac{cz}{H(z)}\right]^{1/3}\,.
\label{eq:D_v}
\end{equation}

Following \cite{Lukovic2016}, we use data from the 6dFGS \cite{Beutler11}, the SDSS DR7 \cite{Ross15}, the BOSS DR11 \cite{Anderson14,Delubac15,Ribera14}, which are reported in Table I of \cite{Lukovic2016}. 
Such a dataset is uncorrelated, therefore  the likelihood can be straightforwardly computed as
\begin{equation}
-2\log{\cal L}_{\mathrm{BAO}}({ \bf p})=\chi _{\mathrm{BAO}}^{2}({ \bf p})= \sum_{i=1}^{6} \left(\frac{\hat\Xi({ \bf p},z_i)-\Xi_{obs}(z_i)}{\sigma_{\Xi}(z_i)}\right)^2\,,
\end{equation} 
where  $\sigma_{\Xi}(z)$ is the error on ${\Xi}(z)$.

\subsection{Gamma Ray Burst} 
 
 We use a dataset of 109 GRB given in \cite{Wei10} which have been already used in other cosmological analysis (see for example \cite{Haridasu2017}). The dataset was compiled using the Amati relation \cite{Amati02,Amati08,Amati09}, and it is formed by 50 GRBs at $z<1.4$ and 59 GRBs  spanning the range of redshift $[0.1, 8.1]$. As it is for SNIa, the observable is the distance modulus, which in case of GRBs is related to peak energy and the bolometric fluence (see for more details \cite{Wei10,Haridasu2017}).
The theoretical counterpart is computed using Eq. \eqref{eq:distmod}, and the $\chi^2$ function is defined as follows
\begin{equation}
-2\log{\cal L}_{\mathrm{GRB}}({ \bf p}) =\chi^2_{\mathrm{GRB}}({ \bf p})=\sum_{i=1}^{109} \left(\frac{\mu_{th}(z_{i},{ \bf p}) - \mu_{obs}(z_{i}) }{\sigma_{\mu}(z_{i})}\right)^2\,.
\end{equation}

\subsection{$T_{CMB}$-redshift relation}

The last dataset is represented by the measurements of the CMB temperature at different redshifts. We use 12 data points obtained by using multi-frequency measurements of the Sunyaev-Zel'dovich effect produced by 813 galaxy clusters stacked on the CMB maps of the {\it Planck} satellite \cite{hurier14}. To those data, we add 10 high redshift measurements obtained through the study of quasar absorption line spectra \cite{bahcall68}. The full dataset includes 22 data points spanning the redshift range $[0.0, 3.0]$, and they are listed in Table I of \cite{Avgoustidis2016}. Finally, we predict the theoretical counterpart using Eq. \eqref{ma3}, and we compute the likelihood as
\begin{equation}
-2\log{\cal L}_{\mathrm{T_{CMB}}}({ \bf p}) =\chi^2_{\mathrm{T_{CMB}}}({ \bf p})=\sum_{i=1}^{22} \left(\frac{T_{CMB, th}(z_{i},{ \bf p}) - T_{CMB, obs}(z_{i})}{\sigma_{T_{CMB}}(z_{i})}\right)^2\,.
\end{equation}

\subsection{PlanckTT+lowP}

The CMB power spectrum is the most powerful tools used to constrain cosmological parameters. However, the calculation of the power spectrum is time consuming, and it is common to use the so called reduced parameters. It is possible to compress the whole information of the CMB power spectrum in a set of four parameters \cite{Kosowsky2002,Wang2007}: the CMB shift parameter ($R$), the angular scale ($l_A$) of the sound horizon at the redshift of the last scattering surface ($z_*$), the baryon density, the scalar
spectral index. Here, we will rely only on $R$ and $l_A$ which can be compute as follows:
\begin{align}
& R = \sqrt{\Omega_{m,0}}\int_0^{z_*}\frac{dz'}{E(z')}\,,\\
& l_A = \frac{\pi D_A(z_*)}{r_s(z_*)}\,,
\end{align}
where $r_s$ is the sound horizon at $z_*$. In the 2015 data release of {\em Planck} satellite, the observational value of those parameters is: 
$[R,l_A]=[1.7488; 301.76]\pm[0.0074; 0.14]$ (for more details see Sect. 5.1.6 in \cite{Planck16}). Thus,  the likelihood can be straightforwardly computed as
\begin{align}
& -2\log{\cal L}_{\mathrm{{CMB}}}({ \bf p}) =\chi^2_{\mathrm{{CMB}}}({ \bf p})= \biggl(\frac{R_{obs}-R_{th}({ \bf p})}{\sigma_R}\biggr)^2+\biggl(\frac{l_{A,obs}-l_{A,th}({ \bf p})}{\sigma_{l_A}}\biggr)^2\,.
\end{align}

\section{Results and discussions}\label{sec:results}

Following the aforementioned methodology, we have carried out two sets of analysis: (A) we fit the whole parameter space composed by the Hubble constant $H_0$, the matter density parameter $\Omega_{m,0}$,  $\gamma$ and $m$; (B) we set $H_0=67.37\pm0.54$ and $\Omega_{m,0}=0.3147\pm0.0074$ which are the best fit values of joint analysis of the CMB power spectrum and other probes \cite{Planck18}, while $m$ and $\gamma$ stay free to vary. All results are summarized in Table \ref{tab:tab2}, and some comments are deserved.
\begin{table}
	\centering
\begin{tabular}{lcccc} 
\hline
	\rm Dataset & \multicolumn{4}{c}{$[H_0,\Omega_{m,0}]$\ free}  \\
	\hline
	& $H_0$ & $\Omega_{m,0}$ & $m$ &  $\gamma$ \\
	\hline
	$H(z)$+$T_{CMB}$ & $66.9^{+2.56}_{-2.34}$ & $0.314^{+0.055}_{-0.045}$ & $0.07^{+0.16}_{-0.14}$ & $1.34\pm0.02$  \\
	SNIa+$H(z)$+$T_{CMB}$ & $71.02^{+0.85}_{-0.91}$ & $0.26\pm 0.03$ & $0.01\pm 0.11$ & $1.34\pm 0.02$   \\
	SNIa+GRB+$H(z)$+$T_{CMB}$ & $71.46^{+0.84}_{-0.85}$ & $0.25\pm{0.03}$ & $0.03^{+0.10}_{-0.11} $& $1.34\pm0.02$ \\
	SNIa+GRB+$H(z)$+BAO+$T_{CMB}$ & $70.31^{+0.66}_{-0.62}$ & $0.30\pm0.01$ & $0.18\pm0.06$ & $1.36\pm0.01$  \\
	SNIa+GRB+$H(z)$+BAO+$T_{CMB}$+CMB & $69.8\pm0.6$ & $0.29\pm0.01$ & $0.01\pm0.02$ & $1.335\pm0.005$   \\
	\hline
	& 
	\multicolumn{4}{c}{$[H_0,\Omega_{m,0}] = [ 67.37,0.315]$ }  \\  
	\hline
	$H(z)$+$T_{CMB}$ & & & $0.08\pm0.07$ & $1.34\pm0.01$ \\
	SNIa+$H(z)$+$T_{CMB}$  & & & $0.05\pm0.07$ &$1.34\pm0.01$ \\
	SNIa+GRB+$H(z)$+$T_{CMB}$   & & & $0.04^{+0.07}_{-0.08}$ & $1.34\pm0.01$ \\
	SNIa+GRB+$H(z)$+BAO+$T_{CMB}$ & & & $0.05\pm0.06$ & $1.339\pm0.009$  \\
	SNIa+GRB+$H(z)$+BAO+$T_{CMB}$+CMB & & & $0.01\pm0.02$ & $1.332\pm0.005$ \\
\hline
\end{tabular}
\end{table}

In the analysis (A), we show that the best fit values of  $[H_0, \Omega_{m,0}]$ are consistent with the most common cosmological analysis at low redshift, and  $[m, \gamma]$ are compatible with the ones from \cite{jetzer10,jetzer11} and their standard values at $1\sigma$ meaning that DE is well described by a cosmological constant. Interestingly, although our parameter space is larger than previous analysis, we get a comparable precision in $m$. This fact expresses the constraining powerful of this dataset with respect to the one used in previous analysis. The matter density is always compatible with current constraint from {\it Planck} 2018 results  \cite{Planck18} at $\sim2\sigma$. Nevertheless, there are two cases in which the central value of $\Omega_{m,0}$ get closer to the one from {\it Planck} at $\sim1\sigma$: (i) when using only $H(z)$ and CMB temperature data; (ii) when using all datasets. Also, the central value of the Hubble constant deserves some comments. In case we used only $H(z)$ and $T_{CMB}$ datasets, we obtain a lower central value of $H_0$ that is compatible at $1\sigma$ with {\it Planck} 2018 constraints and at $3\sigma$ with recent constraint from SNIa \cite{Riess2016,Riess2018}. On the contrary, when introducing luminosity distances measurements, the best fits values of $H_0$ increases showing a tension with 
{\it Planck} 2018 results. The agreement of $H_0$  from the expansion rate data is rather expected since it has been find in other recent analysis \cite{Yu2018,Valent2018,Wang2018}.

Interestingly, the central value of $m$ in the analysis including all the background observables is higher and it is compatible with zero only at $3\sigma$.
In such as case the power law index is $m=0.18\pm0.06$ which can be recast in term of the equation of state parameter using Eq. \eqref{eq:weff} and obtaining $w_{eff}=-0.94\pm0.02$, which is in tension with latest results from Planck satellite ($w=-1.04\pm0.1$ \cite{Planck18}).
This fact demands a deeper analysis to be done with forthcoming datasets such as LSST, Euclid and WFIRST which will explore the Universe until redshift $z\sim 6$ providing  high redshift SNIa and BAO data, and growth factor data with unprecedented precision \cite{lsst, euclid, wfirst, atrio2015}. Finally, in the full analysis including also the CMB constraints we found a lower value of $m$ which can be translated in
$w_{eff}=-0.996\pm0.007$, that is perfectly compatible with a cosmological constant. To compare the decaying DE model with $\Lambda$CMD, we apply the AIC criteria obtaining $\Delta(AIC)=1.53$ which slightly favours the standard cosmological model over the decaying DE one.

In the second analysis, $H_0$ and $\Omega_{m,0}$ are fixed to the {\it Planck} 2018 best fit values, and the parameters $m$ and $\gamma$ are fully in agreement with their expected values. Our best constraint of the power law index is $m=0.01\pm0.02$ which means $w_{eff}=-0.996\pm0.007$ fully compatible with {\it Planck} 2018 results, and with a cosmological constant at $1\sigma$.  
Moreover, to directly compare our results with the ones in \cite{jetzer10,jetzer11}, we carried out another analysis setting $\gamma=4/3$ and leaving only $m$ as free parameter. The constrained values of $m$ with $1\sigma$ error is: $m=0.004\pm 0.006$; which represents a factor of $\sim5$ improvement in $\sigma_m$ over previous constraints. 

\begin{figure}[!ht]
	\centering
	\includegraphics[width=\columnwidth]{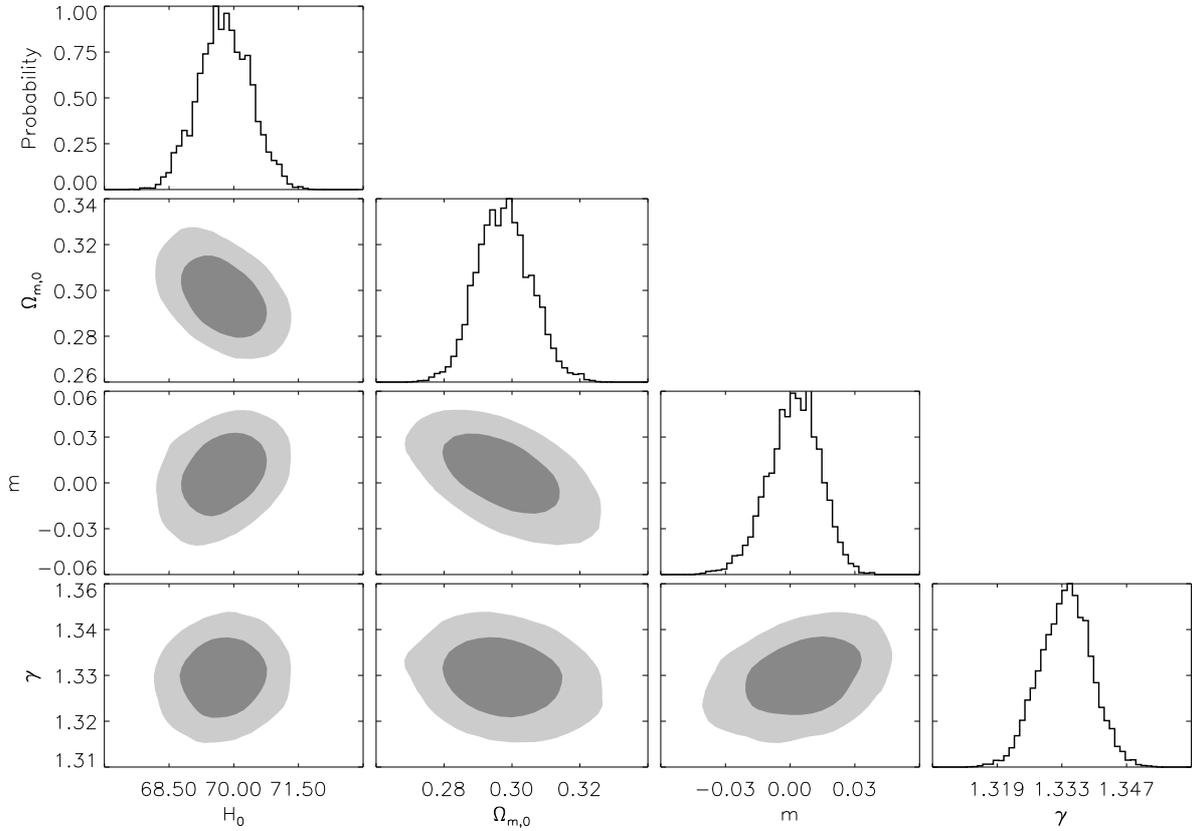}\\	
	\caption{2D marginalized contours of the model parameters $[H_0, \Omega_0, m, \gamma]$  obtained from the MCMC analysis. The 68\% (dark grey) and 95\% (light grey) confidence levels are shown for each pair of parameters. In each row, the marginalized likelihood distribution is also shown.}\label{fig2}
\end{figure}

\begin{figure}[!ht]
	\centering
	\includegraphics[width=\columnwidth]{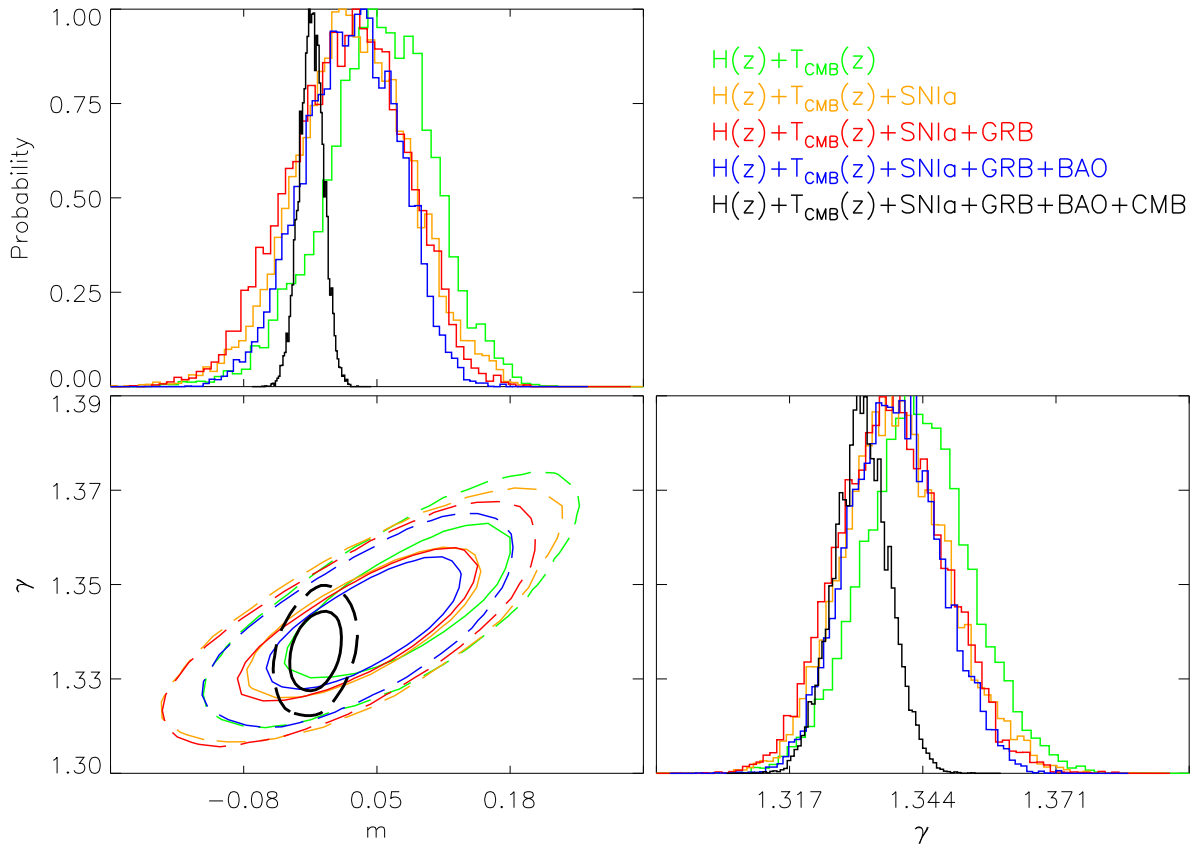}\\	
\caption{2D marginalized 68\% (solid line) and 95\% (dashed line) contours of the model parameters $[ m, \gamma]$  obtained from the MCMC analysis.}\label{fig3}
\end{figure}

Finally, in Fig. \ref{fig2} and  \ref{fig3}, we show the 68\% and 95\% of confidence levels of the whole and reduced parameter space constrained with the full dataset. To avoid overcrowding, in Fig. \ref{fig2}, we do not overplot the contours from the several combinations of the datasets listed in Table \ref{tab:tab2}. 

\section{Conclusions}\label{sec:conclusions}

We have studied the decaying DE model introduced in \cite{ma,jetzer10,jetzer11}. In this model, photons and DM particles can be created or disrupted violating the conservations laws and altering the CMB temperature-redshift scaling relation. The model has been studied using the latest dataset of SNIa, GRB, BAO, $H(z)$, $T_{CMB}(z)$ and $PlanckTT+lowP$ data, that are described in Sect. \ref{sec:data}. 

First, we have explored the whole parameter space composed by the Hubble constant, the matter density fraction, and the parameters $m$ and $\gamma$ introduced in \cite{jetzer10}. In this configuration, when using all the background observables, we obtain that the parameter $m$, that is the power law index of the DE decay law, is compatible with a cosmological constant only at $3\sigma$. Therefore, forthcoming dataset could find a statistically relevant departure from standard cosmology, or alleviate this tension. Nevertheless, it is worth to note that adding the CMB constraints, such a tension disappears.
Second, we have also studied a reduced parameter space composed by only $m$ and $\gamma$, and setting the Hubble constant and the matter density parameter to their best fit values obtained recently by {\it Planck} satellite \cite{Planck18}. In this case, both parameters are always compatible at $1\sigma$ level with standard cosmology. Third, varying only $m$ as in \cite{jetzer10,jetzer11}, we have improved the previous constraints of a factor $\sim 5$.

Finally, on one side we have demonstrated the improved constraining power of current dataset with respect to previous analysis. On the other side, we expect that forthcoming higher precision measurements of the CMB temperature at the location of high redshift galaxy clusters and Quasars, high redshift SNIa, improved measurements of BAO and of luminosity distance of GRBs,
will be able to confirm or rule out decaying DE models \cite{lsst,euclid,wfirst,atrio2015}.

\vspace{6pt}

\funding{ ``This research received no external funding''}

\acknowledgments{
This article is based upon work from COST Action CA1511 Cosmology and Astrophysics Network for Theoretical Advances and Training Actions (CANTATA), supported by COST (European Cooperation in Science and Technology).}

\conflictsofinterest{The authors declare no conflict of interest.} 

\abbreviations{The following abbreviations are used in this manuscript:\\

\noindent 
\begin{tabular}{@{}ll}
BAO  & Baryon Acoustic Oscillation\\
DE   & Dark Energy\\
DM   & Dark Matter\\
FRW  & Friedman-Robertson-Walker\\
GR   & General Relativity\\
GRB  & Gamma Ray Burst\\
MCMC & Monte Carlo Markov Chain\\
SNIa & Supernovae Type Ia\\
SPT  & South Pole Telescope\\
\end{tabular}}

\reftitle{References}

\end{document}